# MODEL-MATCHING PRINCIPLE APPLIED TO THE DESIGN OF AN ARRAY-BASED ALL-NEURAL BINAURAL RENDERING SYSTEM FOR AUDIO TELEPRESENCE

*Yicheng Hsu*[1], *Chenghung Ma*[1], *and Mingsian R. Bai*[1,2]

[1]Department of Power Mechanical Engineering, National Tsing Hua University, Taiwan
[2]Department of Electrical Engineering, National Tsing Hua University, Taiwan

## ABSTRACT

Telepresence aims to create an immersive but virtual experience of the audio and visual scene at the far-end for users at the near-end. In this contribution, we propose an array-based binaural rendering system that converts the array microphone signals into the head-related transfer function (HRTF)-filtered output signals for headphone-rendering. The proposed approach is formulated in light of a model-matching principle (MMP) and is capable of delivering more immersive experience than the conventional localization-beamforming-HRTF filtering (LBH) approach. The MMP-based rendering system can be realized via multichannel inverse filtering (MIF) and multichannel deep filtering (MDF). In this study, we adopted the MDF approach and used the LBH as well as MIF as the baselines. The all-neural system jointly captures the spatial information (spatial rendering), preserves ambient sound (enhancement), and reduces noise (enhancement) before generating binaural outputs. Objective and subjective tests are employed to compare the proposed telepresence system with two baselines.

## 1. INTRODUCTION

Audio Telepresence (AT) refers to technologies that allow one or more users at the near-end to have an immersive experience of the audio scene at the far end. As can be regarded as audio virtual reality, AT aims to transport1the far-end acoustic scene to the near-end user(s) such that the spatial impression of sources and the ambient sounds at the far-end are preserved and reproduced at the near-end. In this context, only mild enhancement should be applied and source separation may not be necessary. Therefore, the AT problem *per se* differs from traditional separation and enhancement problems. At the near-end, either a loudspeaker array or a headphone can be adopted as a means for rendering. In this paper, we focus on binaural reproduction using headphones.

Binaural reproduction of spatial sound enables a spatial audio impression for headphone listeners and finds application in virtual and augmented reality, video conferencing, hearing aids, etc. Interaural Time Difference (ITD) and Interaural Level Difference (ILD) are two spatial cues for localization [1-3]. To produce binaural signals, spherical microphone arrays can be used as the sound capture [4, 5], followed by Head related transfer function (HRTF) filtering [6, 7]. Signal enhancement can be applied to suppress interference or noise [8, 9, 10]. These systems rely on processing the audio signals in the spherical harmonic domain, as frequently used in high-order Ambisonics. A comprehensive review of binaural audio is given by Rafaely et al. [11].

To convert array microphone signals to binaural outputs, a three-stage Localization-Beamforming-HRTF filtering (LBH) approach can be used [12]. In the localization stage, the directions of the sources are determined [13, 14]. In the beamforming stage, beamformers such as the minimum power distortionless response (MPDR) beamformer [15] can be used to extract source signals, based on the direction of arrivals estimated in the localization stage. In the last stage, the beamformer output signals are filtered by the HRTFs associated with the source directions. The performance of the LBH approach relies on localization accuracy and can fail in highly reverberant environments, or when the sources outnumber the microphones.

As opposed to the conventional LBH approaches, we revisit audio telepresence with binaural rendering from the perspective of the model-match principle (MMP). The HRTFs corresponding to 72 virtual source directions uniformly spaced on the horizontal plane are selected in the desired model. This amounts to 5-degree angular spacing between two adjacent sources, which conforms to the minimum audible angle difference of human [16]. By MMP, an overdetermined linear system would be obtained if there are more preselected source directions than microphones, which allows for more degrees of freedom in formulating multichannel inverse filtering (MIF) [17]. In this contribution, we take a step further along the line of MMP from the digital signal processing (DSP)-based approach to a deep learning-based system. We propose a multichannel deep filtering network (MDFnet) implemented using an efficient convolutional recurrent network (CRN) [18] to estimate directly the inverse post-filters. Moderate levels of noise reduction and dereverberation are also incorporated into the proposed MDFnet.

Simulations are performed to validate the proposed learning-based AT system for a six-element uniform circular array (UCA). The conventional LBH and MIF methods are adopted as baselines. ITD and ILD are adopted to quantify the objective performance for spatial impression. In addition, listening tests are also undertaken in terms of subjective indices, perceived directionality, ambience, noise reduction, dereverberation, and artifacts.

This work was supported by the National Science and Technology Council (NSTC) in Taiwan, under the project number 110-22221-E-007-0227-MY3.

## 2. MODEL-MATCHING PRINCIPLE APPLIED TO BINAURAL RENDERING

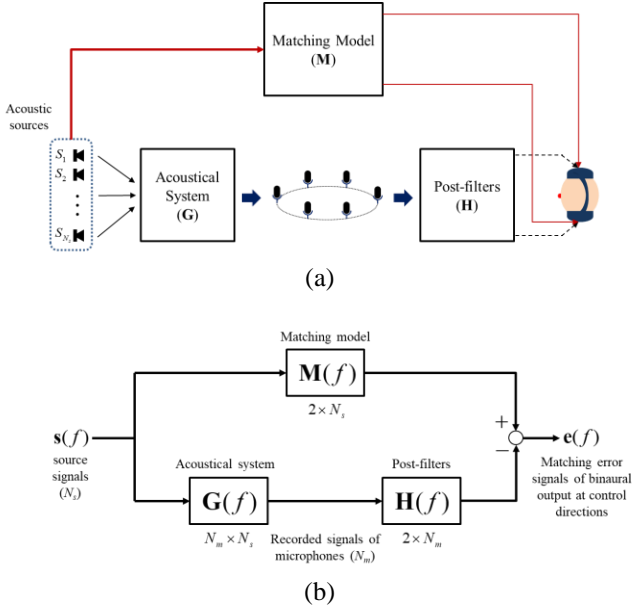

**Fig. 1.** The binaural rendering problem from the model-matching perspective. (a) Binaural recording for headphone rendering, and (b) the general multichannel model-matching framework.

As mentioned above, the AT system investigated in this study performs binaural rendering at the near-end. We may visualize the binaural rendering problem from a model-matching perspective, as illustrated in Fig. 1. The vector $\mathbf{s}(f)$ denotes the input signals from the preselected virtual source directions at the frequency $f$. $\mathbf{M}(f)$ denotes the desired frequency response model. $\mathbf{G}(f)$ denotes the acoustical system represented by acoustic transfer functions. $\mathbf{H}(f)$ denotes the (inverse) post-filters, and $\mathbf{e}(f)$ denotes the matching error vector. The goal of model matching is to match the combined responses of the acoustical system and the post-filters to the desired model (HRTFs). For $N_m$ microphones and $N_s$ preselected source directions, this can be posed as the following optimization problem:

$$\underbrace{\begin{bmatrix} m_{L1}(f) & \cdots & m_{LN_s}(f) \\ m_{R1}(f) & \cdots & m_{RN_s}(f) \end{bmatrix}}_{\mathbf{M}} = \underbrace{\begin{bmatrix} h_{1L}(f) & \cdots & h_{N_mL}(f) \\ h_{1R}(f) & \cdots & h_{N_mR}(f) \end{bmatrix}}_{\mathbf{H}} \underbrace{\begin{bmatrix} g_{11}(f) & \cdots & g_{1N_s}(f) \\ \vdots & \ddots & \vdots \\ g_{N_m1}(f) & \cdots & g_{N_mN_s}(f) \end{bmatrix}}_{\mathbf{G}}, \quad (1)$$

where $g_{ij}$, $i = 1,\ldots,N_m$, $j = 1,\ldots,N_s$, denotes the frequency response function between the $i$-th microphone and the $j$-th virtual source, $h_{iL}$ and $h_{iR}$ denote the post-filters corresponding to the left ear and the right ear, and $m_{Lj}$ and $m_{Rj}$ denote the desired HRTF between the left ear and the right ear for the $j$-th source direction

Often, the number of the preselected virtual source directions is much larger than the number of microphones ($N_s \gg N_m$). MMP leads to an overdetermined system of linear equations. In this case,

a least-square solution of Eq. (1) can be obtained using the Tikhonov regularization [19]:

$$\mathbf{H}(f) = \mathbf{M}(f)\mathbf{G}^H(f)\left[\mathbf{G}(f)\mathbf{G}^H(f) + \beta^2\mathbf{I}\right]^{-1}, \quad (2)$$

where $\beta \in [0,\infty)$ is a regularization parameter. This approach is referred to as the MIF approach hereafter and will be employed as a baseline in the simulation. The rational underlying the MIF approach is significantly different from the LBH approach which requires the source localization, signal extraction, and HRTF filtering steps to be performed in tandem.

## 3. PROPOSED MDF-BASED BINAURAL RENDERING SYSTEN

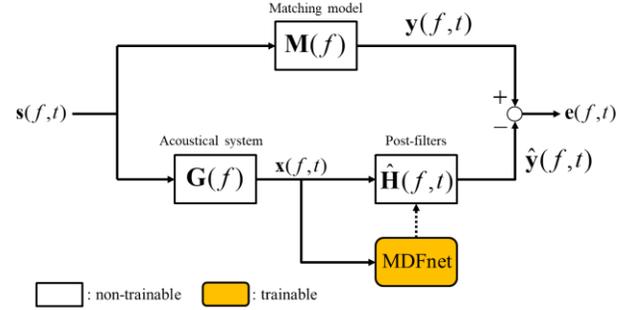

**Fig. 2.** The block diagram of the proposed MDF-based binaural rendering system.

Two limitations of the MIF approach should be mentioned. First, the overdetermined system results in a non-zero residual matching error that cannot be further reduced. Second, noise reduction is not implemented in the MIF design. In this section, a novel approach based on MDF is proposed to address the limitations of the MIF design. The overall system diagram is illustrated in Fig. 2, where $\mathbf{x}(f,t) \in \mathbb{C}^{N_m \times 1}$ denotes the signals captured by microphone array at the $t$-th time frame, and $\hat{\mathbf{H}}(f,t) \in \mathbb{C}^{2 \times N_m}$ denotes the time-varying post-filters that are estimated by the MDFnet as detailed in the sequel. The MDFnet is used to estimate the time-varying post-filters that can transport the far-end acoustic scene to binaural signals for near-end users with mild enhancement.

### 3.1. MDFnet

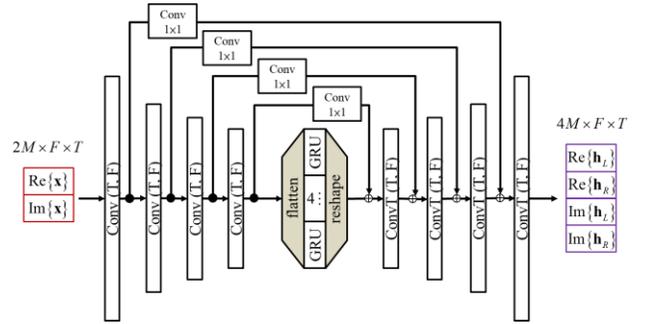

**Fig. 3.** MDFnet architecture with 4 encoder/decoder layers and a bottleneck with 4 parallel recurrent layers.

The MDFnet is based on the CRN and the U-Net [18, 20] as shown in Fig. 3. The MDFnet has four symmetric encoder and decoder layers with a 16-32-64-128 filter. We input the stacked real and imaginary parts of the microphone signals into the encoder, whereas the decoder layer outputs the estimated poster-filters. The convolutional blocks consist of a separable convolution, followed by batch normalization and ReLU activation. In the output layer, a hyperbolic tangent activation function is used to bound the estimated filter coefficients to the range of [-1, 1] [21]. The convolution kernel and step size are set to (3,2) and (2,1). Instead of concatenating the encoder output with the corresponding decoder input, 1×1 pathway convolutions are employed as add-skip connections, which leads to considerable parameter reduction with little performance degradation [18]. The computational complexity of the GRU layer is further reduced by using the grouping technique [18, 22]. The input layer is subdivided into $P$ groups, resulting in $P$ smaller GRU layers. $P = 4$ in the grouped GRU layer.

### 3.2. Training procedure and Loss function

The signal frames were prepared as 32 ms in length with stride 16 ms, where 512-point FFT was used. The optimizer used in training was Adam with a learning rate of 0.001. A gradient norm clipping of 3 is used. The learning rate will be halved if the loss of the validation set fails to improve for three consecutive epochs. We adopted the complex compressed mean-squared error loss [23] that consists of the magnitude-only term and a phase-aware term.

$$L_{spec} = (1-\lambda)\sum_{t,f}\left\||\mathbf{Y}|^c - |\hat{\mathbf{Y}}|^c\right\|_F^2 + \lambda\sum_{t,f}\left\||\mathbf{Y}|^c e^{j\angle\mathbf{Y}} - |\hat{\mathbf{Y}}|^c e^{j\angle\hat{\mathbf{Y}}}\right\|_F^2, \quad (3)$$

where $c = 0.3$ is a compression factor and $\lambda = 0.2$ is a weighting factor. The frequency and time indices are omitted for brevity. The magnitude compression and angle extraction are element-wise operations.

## 4. SIMULATION STUDY

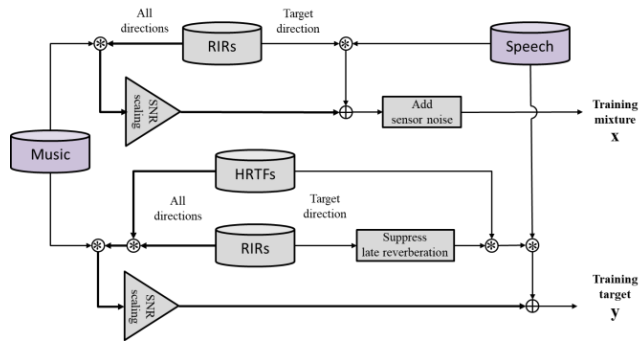

**Fig. 4.** Preparation of the training data of the mixture and target signals.

### 4.1. Data preparation

Clean speech signals consisting of utterances from 921 and 40 speakers were selected from the *train-clean-360* and *dev-clean* subsets of the LibriSpeech corpus [24] for training and testing of the proposed network. Music signals from Free Music Archive [25] were used to simulate the ambient sounds. The simulation was conducted at a sample rate of 16 kHz. The desired model (**M**) was based on the HRTF database provided by York University [26]. The room impulse responses (RIRs) of the acoustic system (**G**) at 72 uniformly spaced source angles in the horizontal plane were simulated by using the image source method [27] with reverberation time (T60) = 0.2, 0.4, and 0.6 s. A six-element UCA with 10 cm diameter was used. The data preparation for training and validation is illustrated in Fig. 4. Music signals played at 72 uniformly spaced directions serve as ambient sounds. Five-second clips of noisy signals were prepared by mixing a target speech signal and the ambient sounds with signal-to-ambience ratio (SAR) = 0, 5, 10, 15, and 40 dB. In addition, sensor noise was added with signal-to-noise ratio (SNR) = 20, 25, and 30 dB. To ensure natural-sounding target speech signals, only moderate dereverberation was applied as in [18].

### 4.2. Implementation of the baselines and proposed approach

Two baseline approaches were employed to benchmark the proposed MDF-based AT system. One baseline was the LBH approach which performs source localization, signal extraction, and HRTF filtering in tandem. Steered-response power phase transform and MPDR were used in the first two stages, where free-field steering vectors in both stages were used. Another baseline was the MIF approach with the regularization parameter $\beta = 0.0001$. The acoustic system matrix **G** for MIF was simulated using ISM at 72 angles in the horizontal plane.

### 4.3. Results

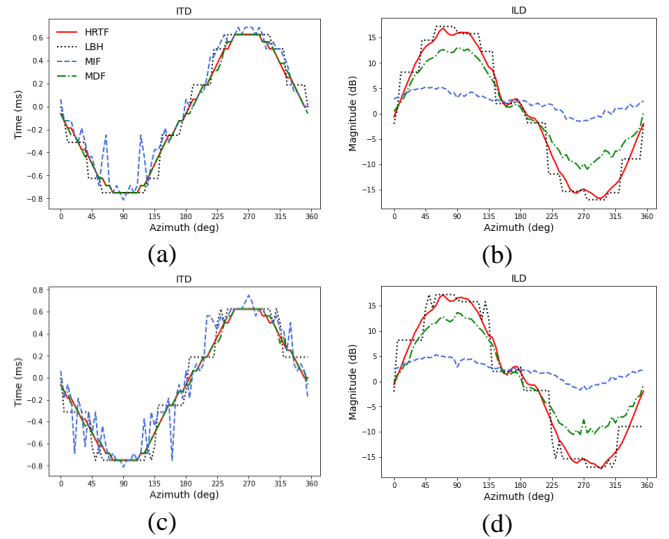

**Fig. 5.** The comparison of ITD and ILD with SNR = 30 dB, and T60 equals (a)(b) 320 ms, and (c)(d) 540 ms.

We compared the proposed MDF approach with the two baselines by means of the objective and subjective tests. In the objective test, we used ITD and ILD as the performance measures for localization of a target speaker. ITD was calculated up to 1500 Hz as the low-frequency spatial cue [28]. The results in Fig. 5 suggest that the

LBH can fit the ITD and ILD curves relatively well in mild reverberation condition (T60 = 320 ms). However, it can be seen from the result that LBH produces noisier ITD and ILD curves than MDF at a higher reverberation level (T60 = 540 ms). Furthermore, MIF performed the worst in ILD due to the non-zero residual matching error of the overdetermined system. Therefore, the ITD and ILD of binaural signals obtained using the proposed MDF approach remain reliably rendered in reverberation environments, which is a highly desirable feature of MDF for real-world applications.

To further compare three audio telepresence approaches, we define an objective measure to quantify the rendering performance. The audio telepresence matching error ($E_{ATM}$) is defined as the difference between the reproduced binaural signals, $\hat{\mathbf{Y}}(f,t)$, and the desired signals, $\mathbf{Y}(f,t)$, as follows:

$$E_{ATM} = 20\log_{10}\left(\frac{1}{FT}\sum_{f,t}\left\|\mathbf{Y}(f,t)-\hat{\mathbf{Y}}(f,t)\right\|_F\right), \quad (4)$$

where $F$ and $T$ denote the number of frequency bins and time frames, respectively. $E_{ATM}$ calculated for two reverberation conditions are summarized in Table 1. The MDF approach performs the best among three approaches. Furthermore, two MMP-based methods are more robust to reverberation than the LBH method that could suffer from performance degradation due to localization error in high reverberation conditions.

**Table 1.** The audio telepresence matching error calculated for T60 = 320 ms and 540 ms. The best performing cases are boldfaced.

| T60 (ms) | LBH | MIF | MDF |
|---|---|---|---|
| 320 | -34.70 | -32.91 | **-40.35** |
| 540 | -32.58 | -32.80 | **-40.82** |

In the following, we consider simulation scenario, where a target speaker source moves in a circle with SAR = 15 dB and SNR = 25 dB, as depicted in Fig. 6. Reverberation is simulated with T60 = 540 ms. The subjective listening test was conducted according to the multiple stimuli with hidden reference and anchor (MUSHRA) procedure [29]. The training target signals in Fig. 4 and highpass filtered signals were used as the reference and the anchor. Sense of direction, ambience, noise reduction, dereverberation, and artifacts were adopted as the subjective measures. The subjective score ranged from 1 to 5. In total, 15 listeners of different genders, experts, and non-experts participated in the listening test. The MUSHRA results are presented in Fig. 7. The proposed MDF approach significantly outperformed the two baselines in overall performance. Although LBH provided a very good sense of direction, it failed to preserve the ambient sounds, resulting in a poor sense of ambience compared to the model-matching approaches. Limited by the number of microphones, the MIF method could not reproduce the directional target speaker well. In contrast, if the model matching AT system is implemented using the learning-based approach, the performance improvement is quite dramatic. In addition, MIF was ineffective in reducing noise and reverberation. In summary, the overall AT performance of the three approaches can be ranked as MDF, MIF, and LBH. To assess the processing speed, Real Time Factor (RTF) is calculated using the CPU, i9-12900@2.4GHz in a single thread for all approaches. The smaller the RTF is, the faster the algorithm runs.

The RTFs of LBH, MIF, and MDF turn out to be 0.853, 0.017, and 0.160. The two MMP-based approaches run much faster than the LBH approach.

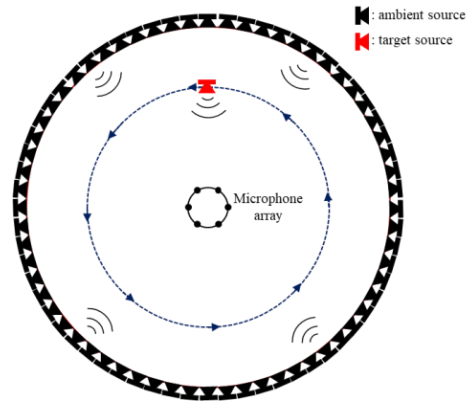

**Fig. 6.** Simulation settings. The microphone array, the moving target source, and the ambient sources are indicated.

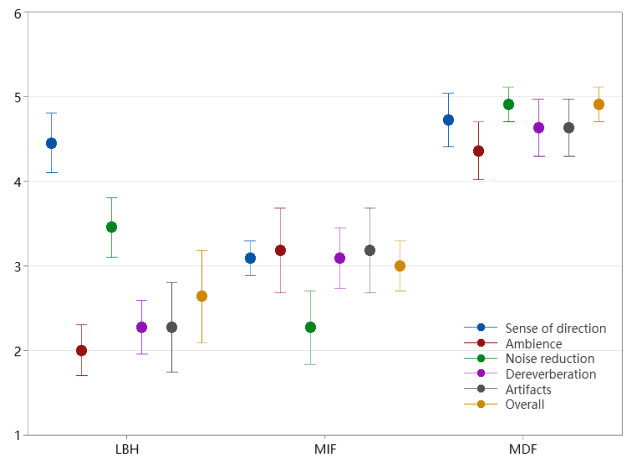

**Fig. 7.** The ANOVA output of the listening test for a moving speech source from 0º to 360º and ambient sound.

## 5. CONCLUSIONS

A novel learning-based AT system with binaural rendering is proposed in this paper. The contributions can be summarized as follows. First, the AT problem and the associated performance measure are defined. Second, the AT system is designed within a generalized MMP framework. This one-stage approach is far more computationally efficient than the conventional LBH approach. Third, in light of MMP, an MDF network that allows for superior AT performance with minor computational cost is implemented via a learning-based approach. In conclusion, the MDF approach can preserve the spatial impression of the target source and the ambient sounds, and suppress excessive noise and reverberation in AT applications.

# 6. REFERENCES

[1] J. C. Middlebrooks, J. C. Makous, and D. M. Green, "Directional sensitivity of sound-pressure levels in the human ear canal," *J. Acoust. Soc. Amer.*, vol. 86, no. 1, pp. 89-108, Jul. 1989.

[2] B. G. Shinn-Cunningham, S. G. Santarelli, and N. Kopco, "Tori of confusion: Binaural cues for sources within reach of a listener," *J. Acoust. Soc. Amer.*, vol. 107, no. 3, pp. 1627-1636, 2000.

[3] J. Braasch, "Modelling of binaural hearing," in *Communication Acoustics*, New York: Springer, pp. 75-108, 2005.

[4] T. D. Abhayapala and D. B. Ward, "Theory and design of high order sound field microphones using spherical microphone array," in *Proc. IEEE ICASSP*, 2002, vol. 2, pp. 1949-1952.

[5] D. L. Alon, J. Sheaffer, and B. Rafaely, "Robust plane-wave decomposition of spherical microphone array recordings for binaural sound reproduction," *J. Acoust. Soc. Amer.*, vol. 138, no. 3, pp. 1925-1926, 2015.

[6] W. Zhang, T. D. Abhayapala, R. A. Kennedy, and R. Duraiswami, "Insights into head-related transfer function: Spatial dimensionality and continuous representation," *J. Acoust. Soc. Am.*, vol. 127, bo. 4, pp. 2347-2357, 2010.

[7] M. Jeffet, N. R. Shabtai, and B. Rafaely, "Theory and perceptual ecaluation of the binaural reproduction and beamforming trade-off in the generalized spherical array beamformer," *IEEE/ACM Trans, Audio, Speech, Lang. Proc.*, vol. 24, no. 4, pp. 708-718, 2016.

[8] C. Borrelli, A. Canclini, F. Antonacci, A. Sarti, and S. Tubaro, "A denoising methodology for higher order ambisonics recordings," in *2018 16th Intl. Workshop Acoust. Signal Enhancement (IWAENC)*, pp. 451-455, 2018.

[9] M. Lugasi and B. Rafaely, "Speech enhancement using masking for binaural reproduction fo ambisonics signals," *IEEE/ACM Trans, Audio, Speech, Lang. Proc.*, vol. 28, pp. 1767-1777, 2020.

[10] A. Herzog and E. A. Habets, "Direction and reverberation preserving noise reduction of Ambisonics signals," *IEEE/ACM Trans, Audio, Speech, Lang. Proc.*, vol. 28, pp. 2461-2475, 2020.

[11] B. Rafaely et al., "Spatial audio signal processing for binaural reproduction of recorded acoustic scenes—review and challenges," *Acta Acustica*, vol. 6, no. 47, 2022.

[12] H. Beit-On et al., "Audio signal processing for telepresence based on wearable array in noisy and dynamic scenes," in *Proc. IEEE ICASSP*, 2022, pp. 8797-8801.

[13] Z. Khan, M. M. Kamal, N. Hamzah, K. Othman, and N. Khan, "Analysis of performance for multiple signal classification (MUSIC) in estimating direction of arrival," in *2008 IEEE Intl. rf and Microwave Conf.*, 2008, pp. 524-529.

[14] V. Tourbabin, D. L. Alon, and R. Mehra, "Space domain-based selection of direct-sound bins in the context of improved robustness to reverberation in direction of arrival estimation," in *Proc. 11th European Cong. And Exposition on Noise Control Engineering*, 2018, pp. 2589-2596.

[15] H. L. V. Trees, Optimum Array Processing, Detection, Estimation, and Modulation Theory, Part IV,. New York: Wiley, 2002.

[16] C. Freigang, K. Schmiedchen, I. Nitsche, and R. Rübsamen, "Free-field study on auditory localization and discrimination performance in older adults," *Experimental Brain Research*, vol. 232, no. 4, pp. 1157-1172, 2014.

[17] M. Bertero, T. A. Poggio, and V. Torre, "Ill-posed problems in early vision," in *Proc. IEEE*, vol. 76, no. 8, pp. 869-889, 1988.

[18] S. Braun, H. Gamper, C. K. A. Reddy, and I. Tashev, "Towards efficient models for real-time deep noise suppression," in *Proc. ICASSP*, 2021, pp. 656-660.

[19] A. N. Tikhonov, A. Goncharsky, V. Stepanoc, and A. G. Yagola, Numerical methods for the solution of ill-posed problems, Norwell, MA, USA: Kluwer, 1995.

[20] K. Tan and D. Wang, "A convolutional recurrent neural network for real-time speech enhancement," in *Proc. Interspeech*, 2018, pp. 3229-3233.

[21] W. Mack and E. A. P. Habets, "Deep filtering: Signal extraction and reconstruction using complex time-frequency filters," *IEEE Signal Processing Letters*, vol. 27, pp. 61-65, 2020.

[22] K. Tan and D. Wang, "Learning complex spectral mapping with gated convolutional recurrent networks for monaural speech enhancement," *IEEE/ACM Transactions on Audio, Speech, and Language Processing*, vol. 28, pp. 380–390, 2019.

[23] A. Ephrat, I. Mosseri, O. Lang, T. Dekel, K. Wilson, A. Hassidim, W. T. Freeman, and M. Rubinstein, "Looking to listen at the cocktail party: A speaker-independent audio-visual model for speech separation," *ACM Trans. Graph.*, vol. 37, no. 4, 2018.

[24] V. Panayotov, G. Chen, D. Povey, and S. Khudanpur, "Librispeech: an ASR corpus based on public domain audio books," in *Proc. IEEE ICASSP*, 2015, pp. 5206–5210.

[25] M. Defferrard, K. Benzi, P. Vandergheynst, and X. Bresson, "FMA: A dataset for music analysis," in *Proc. Int. Society for Music Information Retrieval Conf.*, Suzhou, China, pp. 316-323, 2017.

[26] C. Armstrong, L. Thresh, D. Murphy, and G. Kearney, "A perceptual evaluation of individual and non-individual HRTFs: a case study of the SADIE II database," *Applied Sciences*, vol. 8, no. 11, 2018.

[27] E. Lehmann and A. Johansson, "Prediction of energy decay in room impulse responses simulated with an image-source model," *J. Acoust. Soc. Amer.*, vol. 124, no. 1, pp. 269–277, Jul. 2008.

[28] S. D. Ewert, K. Kaiser, L. Kernschmidt, and L. Wiegrebe, "Perceptual sensitivity to high-frequency interaural time differences created by rustling sounds," *J. Assoc. Res. Otolaryngol.*, vol. 13, pp. 131-143, 2012.

[29] ITU-R Recommendation, "Method for the subjective assessment of intermediate sound quality (MUSHRA)," *International Telecommunications Union*, BS. 1534-1, Geneva, Switzerland, 2001.